\documentstyle[sprocl,psfig]{article}

 \def\be{\begin{equation}}
 \def\ee{\end{equation}}
 \def\bea{\begin{eqnarray}}
 \def\eea{\end{eqnarray}}

\begin{document}
 
 \title{DARK MATTER PREDICTIONS WITH NON-UNIVERSAL SOFT BREAKING
  MASSES}

 \author{R. ARNOWITT}
 
 \address{Center for Theoretical Physics, Department of Physics,
 Texas A\&M University, College Station,\\
 TX 77843-4242, USA\\E-mail: arnowitt@phys.tamu.edu}
 
 \maketitle\abstracts{ 
The effect of non-universal SUSY soft breaking on 
predictions of dark matter detector event rates are surveyed for supergravity 
models with gravity mediated soft breaking.
For universal soft breaking in the first two generations and 
$\tan \beta {\tiny \begin{array}{l}  <\\ \sim \end{array}} 20$, non-universal 
effects can be characterized by four parameters (two for Higgs and two for third 
generation squarks) in addition to those of the minimal model (MSGM). 
These can increase or decrease event rates by a factor of 10-100 in the 
domain 
$m_{\chi_1^0}{\tiny \begin{array}{l}  <\\ \sim \end{array}} 60$ $GeV$ 
($\chi_1^0=$ lightest neutralino) but produce generally small effects at higher 
masses.
The value of the top mass and $b\rightarrow s+\gamma$ branching ratio eliminates 
most of the parameter space for $\mu<0$, causing event rates for $\mu<0$ 
 to be a factor $\approx 100$ smaller 
than for $\mu>0$. 
A correlation between large (small) event rates and small (large) 
$b\rightarrow s+\gamma$ branching ratio is observed.
The effect of future satellite (MAP and PLANCK) precision determinations of 
cosmological parameters on predicted event rates is examined for examples of 
the $\Lambda$CDM and $\nu$CDM models. 
It is seen that these could sharpen the event rate predictions and also restrict 
the allowed gaugino mass ranges, thus influencing predictions for accelerator 
SUSY searches.
 }
 
 \section{Introduction}
 
 The composition of the dark matter (DM) in the universe, which makes up
 90\% or more of the universe's total matter, is one of the most important 
 unresolved problems in astronomy.
At present, dark matter has been observed only from its gravitational interactions 
and may consist of a number of different components \cite{jed}.
There may be baryonic (B) dark matter (e.g. machos), hot dark matter (HDM) which 
was relativistic at the time galaxy formation (possibly massive neutrinos) and 
cold dark matter (CDM) which was non-relativistic during galaxy formation.
In  addition a cosmological constant ($\Lambda$) may be present. 

The amount of each type of DM can be specified by the parameter 
$\Omega_i=\rho_i/\rho_c$ where $\rho_i$ is the matter density of type ``$i$'', 
$\rho_c= 3H^2/8\pi G_N$ is the critical matter density to close the universe, $H$ is 
the Hubble constant, $H= h 100km/s Mpc$ and $G_N$ the Newtonian constant.
Currently, measurements of $h$ fall in the range
\begin{equation}
0.5 {\tiny \begin{array}{l}  <\\ \sim \end{array}} h 
{\tiny \begin{array}{l}  <\\ \sim \end{array}} 0.75
\end{equation}
and different models give estimates for the cosmological CDM of
\begin{equation}
0.1\le \Omega_{CDM}h^2 \le 0.4
\end{equation}
The CDM that can be directly observed is that which exists locally in the Milky Way. 
This has been estimated to have a density of $\rho^{MW}_{DM}\simeq 0.3$ $GeV/cm^3$, 
impinging on the Solar System with velocity $v_{DM}\simeq 300 km/s$.

The cold dark matter is of particular interest in that if it is of particle nature it is 
most likely an ``exotic'' particle, i.e. one not found in the Standard Model (SM). 
We consider here models of physics beyond the Standard Model based on supergravity 
grand unification with R-parity invariance, where supersymmetry (SUSY) breaking 
takes place at a scale  near or above the GUT scale $M_G\cong 1.5\times 10^{16}$ $GeV$. 
This breaking occurs in a ``hidden'' sector and is transmitted (super)gravitationally to 
the physical sector \cite{dva}. 
Such models automatically predict the existance of CDM in that the lightest 
supersymmetric particle (LSP) is absolutely stable and over most of the parameter space is the lightest neutralino, $\chi_1^0$.
Thus the relic  $\chi_1^0$ left over from the Big Bang would be the CDM seen today.
Further, over a significant part of the SUSY parameter space, the amount of CDM 
predicted is in accord with what is observed astronomically, i.e. Eq.(2).

 \section{DM Detector Event Rates}

Terrestial experiments detect incident local (Milky Way) DM particles by their 
scattering by quarks in nuclear targets. 
We briefly review in this section the relevant formulae for prediction 
of detector event rates for SUSY models.

The analysis proceeds as follows. 
One first calculates the relic density  of neutralino CDM which remain after 
neutralino annihilation in the early universe.
One finds \cite{jed,tri}: 
\begin{equation}
\Omega_{\chi_1^0}h^2 \cong 2.48\times 10^{-11} 
\biggl ( {T_{\chi_1^0}\over T_\gamma} \biggr )^3 
 \biggl ( { T_\gamma \over 2.73} \biggr )^3
{N_f^{1/2}\over J(x_f)} 
\end{equation}
where $x_f=kT_f/m_{\chi_1^0}$ ($T_f=$ neutralino freezeout temperature, 
$N_f=$ number of degrees of freedom at freezeout), $T_\gamma$ is the cosmic 
microwave background (CMB) temperature, $(T_{\chi_1^0}/T_\gamma)^3$ is the 
reheating factor and 
\begin{equation}
J(x_f)= \int_0^{x_f} dx \langle \sigma v \rangle (x)\ GeV^{-2}
 \end{equation}
In Eq.(4), $\sigma$ is the neutralino annihilation cross section (calculated from 
the SUSY model), $v$ the relative velocity and $\langle \quad \rangle$ means 
thermal average.
One sees that $\Omega_{\chi_1^0}h^2$ scales inversely with the annihilation cross 
section, i.e. the more annihilation there is, the less relic $\chi_1^0$ remain. 
One restricts the SUSY parameter space so that Eq.(2) is obeyed, and also that the 
current SUSY parameter space bounds from LEP, the Tevatron and CLEO 
($b\rightarrow s+\gamma$ decay) are satisfied.

One then calculates, in the restricted SUSY parameter space, the quantity $R$, 
the expected event rate of detector scattering events (per kilogram of detector 
per day) for the incident flux of Milky Way neutralinos \cite{jed}:
\begin{equation}
R= (R_{SI}+R_{SD})\biggl [ {\rho_{\chi_1^0} \over 0.3\ GeV/cm^3}\biggr ]
   \biggl [ {v_{\chi_1^0} \over 320km/s}\biggr ] {events\over kg\ d}
\end{equation}
 where
\begin{equation}
R_{SI}= { 16m_{\chi_1^0} M_N^3 M_Z^4\over [M_N+m_{\chi_1^0}]^2}
\mid A_{SI} \mid ^2 
\end{equation}
\begin{equation}
R_{SD}= { 16m_{\chi_1^0} M_N\over [M_N+m_{\chi_1^0}]^2} 
\lambda^2 J(J+1) \mid A_{SD} \mid ^2 
\end{equation}
where $M_N$ is the nuclear target mass and $J$ is its spin, $M_Z$ is the $Z$ boson 
mass and $\langle N|\sum \vec{S_i} |N \rangle = \lambda \langle N| \vec{J} |N\rangle$
with $\vec{S_i}$ the ith nucleon's spin.
In Eqs.(6) and (7), $A_{SI} (A_{SD})$ are the spin independent (spin dependent) 
scattering amplitudes.
Note that for heavy nuclear targets, $R_{SI}\sim M_N$ while $R_{SD}\sim 1/M_N$ 
making the heavier targets generally more sensitive than lighter ones and $R_{SI}$ 
generally the dominant contribution for heavy targets.
There are a number of uncertainties in the above analysis involving the strange quark 
contribution to the nucleon, the nature of nuclear form factors etc., making the theoretical predictions of $R$ uncertain to perhaps a factor $\approx 2$.

\section{Soft Breaking: Universal Case}

We consider here models where the GUT group $G$ breaks to the Standard Model 
group at $M_G: G\rightarrow SU(3)\times SU(2)\times U(1)$.
The simplest supergravity model of this type, the minimal SUGRA model 
(MSGM) depends at $M_G$ on four extra soft breaking parameters and one sign 
to determine all the masses and interactions of the 32 SUSY particles \cite{dva,cet}. 
These new parameters are $m_0$ (the universal scalar soft breaking mass), 
$m_{1/2}$ (the universal gaugino mass),  $A_0$ (the universal 
cubic soft breaking parameter), $B_0$ (the quadratic soft breaking parameter) and 
the sign of $\mu_0$, the Higgs mixing parameter in the effective potential term 
$\mu_0 H_1 H_2$ (where $H_i, i=1,2$ are the two Higgs doublets). 
The renormalization group equations (RGE) then allow one to proceed downward 
to lower energy scales where the soft breaking parameters trigger the breaking of 
$SU(2)\times U(1)$ \cite{pet}. 
In fact one may show that a necessary condition that electroweak breaking occur at 
a lower scale is that at least one soft breaking parameter and $\mu_0$ be 
non-zero at $M_G$, and this breaking will occur at the electroweak scale ($M_Z$) 
provided $m_t$ obeys 
$90\ GeV {\tiny \begin{array}{l}  <\\ \sim \end{array}} m_t 
{\tiny \begin{array}{l}  <\\ \sim \end{array}}  200$ $GeV $. 
Thus the model automatically requires a heavy top quark and it is SUSY soft breaking 
at $M_G$ that gives rise to electroweak breaking at $M_Z$.

The conditions for electroweak breaking are 
\begin{equation}
\mu^2= {\mu_1^2-\mu_2^2\tan^2\beta\over \tan\beta^2 -1} -{1\over2}M_Z^2;
\quad
\sin^2\beta= -{2B\mu\over \mu_1^2+\mu_2^2+2\mu^2}
\end{equation}
where $\mu^2$, $\mu_i^2$, $B$ are running parameters at $Q=M_Z$, 
$
\mu_i^2= m_{H_i}^2+\sum_i, \quad \tan\beta= 
\langle H_2\rangle /\langle H_1\rangle
$
and $\sum_i$ are loop corrections \cite{ses}.
Thus radiative breaking allows one to determine $\mu^2$ and eliminate $B$ in terms 
of $\tan\beta$. 
The determination of $\mid \mu\mid$ greatly enhances the predictive power of 
the model.
In the following we will restrict the parameter space examined to be in the domain 
\begin{equation}
m_0,m_{\tilde{g}} \le 1 TeV, \quad 2\le\tan\beta\le 25, \quad -7\le A_t/m_0\le 7
\end{equation}
where $m_{\tilde{g}}$ is the gluino mass and $A_t$ is the $t$-quark $A$-parameter 
at the electroweak scale.
(Eq.(9) satisfies usual ``naturalness'' conditions.)
For most of the parameter space, $\mu^2/M_Z^2\gg 1$ which leads to ``scaling'' 
relations \cite{sed} for the charginos ($\chi_i^\pm, i=1,2$) and neutralinos 
($\chi_i^0, i=1,..,4$):
\begin{equation}
2m_{\chi_1^0}\cong m_{\chi_1^\pm} \cong m_{\chi_2^0}\cong 
({1\over3}-{1\over4}) m_{\tilde{g}}; \quad 
m_{\tilde{g}}\cong {\alpha_3(M_Z)\over \alpha_G} m_{1/2}
\end{equation}
\begin{equation}
m_{\chi_2^\pm}\cong m_{\chi_3^0 }\cong m_{\chi_4^0}\gg m_{\chi_1^0}
\end{equation}
In addition, the four Higgs bosons, $h$, $H^0$, $A^0$ and $H^\pm$ obey in this domain 
the ralations 
\begin{equation}
m_{H^0}\cong  m_A \cong m_{H^\pm} \gg m_h
\end{equation}
and $m_h {\tiny \begin{array}{l}  <\\ \sim \end{array}}  120$ $GeV$. 

Most of the SUGRA dark matter analysis has been done within the above framework 
of universal soft breaking \cite{osa,jed}. 
We examine next what modifications arise when non-universal soft breaking occurs.

 \section{Soft Breaking: Non-Universal Case}

The MSGM model with universal soft breaking parameters, is of course the simplest 
SUGRA model, which in part is why it has been examined so extensively.
However, there are a number of reasons why one might expect non-universal 
SUSY soft breaking to occur at $M_G$.
Thus in general,  non-universal soft breaking will arise if in the Kahler potential, 
the interactions between the hidden sector  fields (whose  VEVs  give rise to 
SUSY breaking) and the physical sector fields are not universal \cite{dev}. 
Further, even if universality where to hold at a more fundamental level, e.g. at the 
Planck or string scales, running the RGE down to the GUT scale can produce 
significant non-universal soft breaking at $M_G$ \cite{des}. 
Finally, we note that in the breaking of a higher rank GUT group down to the SM 
group at $M_G$, the $D$ terms can generate non-universalities \cite{11}.

Flavor changing neutral currents [FCNC] will be suppressed if the soft breaking 
masses of the first two generations are universal at $M_G$ \cite{12}, 
and in the following we will assume that this is the case and let $m_0$  be their  
common mass.
We then parametrize the Higgs and third generation squark and slepton masses at 
$M_G$ as follows:
\begin{equation}
m_{H_1}^2 = m_0^2 (1+\delta_1); \quad
m_{H_2}^2 = m_0^2 (1+\delta_2)
\end{equation}
\begin{equation}
m_{q_L}^2 = m_0^2 (1+\delta_3); \quad
m_{u_R}^2 = m_0^2 (1+\delta_4); \quad
m_{e_R}^2 = m_0^2 (1+\delta_5)
\end{equation}
\begin{equation}
m_{d_R}^2 = m_0^2 (1+\delta_6); \quad
m_{l_L}^2 = m_0^2 (1+\delta_7)
\end{equation}
where $q_L=(\tilde{u}_L,\tilde{d}_L)$ is the left squark doublet, 
$l_l=(\tilde{\nu}_L,\tilde{e}_L)$ the left slepton doublet etc.
The $\delta_i$ represent the deviations from universality, and we will restrict 
these to obey $-1\le \delta_i\le +1$.
In addition, we denote the $A$ parameters at $M_G$ by $A_{0t}$, $A_{0b}$ 
and $A_{0\tau}$ which also need not be universal.

For GUT groups containing an $SU(5)$ subgroup with matter embedded in the 
$10+\bar{5}$ of $SU(5)$ (e.g. $SU(N), N\ge 5;\ SO(N), N\ge 10;\ E_6$) one has 
\begin{equation}
\delta_3= \delta_4= \delta_5; \quad \delta_6= \delta_7; \quad 
A_{0b} = A_{0\tau}
\end{equation} 
We note that for $\tan\beta {\tiny \begin{array}{l}  <\\ \sim \end{array}} 20$, 
$\delta_5, \delta_6, \delta_7$ and  $A_{0b}, A_{0\tau}$ do not enter significantly 
in the calculations, and we will set them to zero in the following. 

To examine the significance of the non-universal soft breaking, we exhibit 
the 1-loop RGE expressions for $\mu^2$ of Eq.(8) for $Q=M_Z$ \cite{13}:
\bea
\mu^2 & = & {t^2\over t^2-1} \biggl[ 
    \biggl \{  {1-3D_0 \over 2} + {1\over t^2}  \biggr \} 
\nonumber \\
& & + \biggl \{ {1-D_0 \over 2}(\delta_3+\delta_4) -  {1+D_0 \over 2} \delta_2 
      + {1\over t^2}\delta_1  \biggr \} \biggr ] m_0^2 
\nonumber \\
 & & +  {t^2\over t^2-1}\biggl[ {1\over2}(1-D_0){A_R^2\over D_0} 
 + C_\mu m_{1/2}^2 \biggr ]  - {1\over 2}M_Z^2 
\nonumber \\
 &  &  + {1\over 22}{t^2+1\over t^2-1} 
\biggl( 1-{\alpha_1(M_z)\over \alpha_G}\biggr)S_0
 \label{eq:sp}
 \eea
where $t\equiv \tan\beta$, $D_0\cong 1-(m_t/200\sin\beta)^2$, 
$A_R\cong A_t-0.61 m_{\tilde{g}}$, $\alpha_G \cong 1/24$, $S_0=Tr Y m^2$ 
($Y=$ hypercharge and $m$ are the soft breaking masses at $M_G$) and 
$C_\mu = {1\over2}D_0(1-D_0)(H_3/F)^2+e-g/t^2$ (with the RGE form 
factors $H$, $F$, $e$, $g$ given in Iba\~nez et al \cite{pet}).
In Eq.(17), the $m_0^2$ term has been divided into a universal part and a 
non-universal part (dependent on the $\delta_i$). 
$D_0$ vanishes at the $t$-quark Landau pole and is generally small 
($0<D_0 {\tiny \begin{array}{l}  <\\ \sim \end{array}} 0.25$). 
$A_R$ is the residue at the Landau pole (i.e. $A_{0t}=A_R/D_0-(H_3/F)m_{1/2}$). 
Since $m_{1/2}\cong (\alpha_G/\alpha_3)m_{\tilde{g}}$, where $\alpha_G$ is the 
GUT coupling constant, the $C_\mu m_{1/2}^2$ term scales with $m_{\tilde{g}}^2$ 
or alternately by Eq.(10) with $(m_{\chi_1^0})^2$. 
$S_0$ vanishes for universal soft breaking masses (by hypercharge anomaly  
cancelation) but is non-zero for non-universal masses. 
(The $S_0$ term is generally small, usually only a few percent correction.) 

\section{Effects Of Non-Universal Soft Breaking On DM Event Rates}

While many parameters enter into the DM event rate formula Eq.(5), $\mu^2$ 
plays a central role in determining $R$.
This is because $\mu^2$ governs the interference between the Higgsino and 
gaugino parts of $\chi_1^0$ in the  $\chi_1^0$ - quark scattering amplitude and 
it is this interference which gives rise to $R_{SI}$ (which we have seen for most 
detectors is the dominant part of $R$). 
One finds in general that increasing (decreasing) $\mu^2$ decreases (increases) 
the amount of interference and hence the size of $R$.

Since $D_0$ is small, one sees from Eq.(17) that $\delta_3$ and $\delta_4$ 
contribute oppositely to $\delta_2$ in $\mu^2$.
Thus in evaluating the contribution of non-universal masses, it is necessary 
to consider both Higgs and squark masses since they produce effects  of the 
same size. 
There are certain situations where the non-universal terms become relatively 
large compared to the universal contributions, greatly enhancing the 
non-universal effects. 
Thus  the  
universal contribution of the Landau pole term (which is generally quite large) 
vanishes if the residue $A_R$ is zero, i.e. when $A_t\cong 0.61 m_{\tilde{g}}$. 
Also the $C_{\mu} m_{1/2}$ term is small for light  $\chi_1^0$ (or light 
$\tilde{g}$) according to the scaling relations Eq.(10). 
In addition,  for 
$t^2\gg 1$ (i.e. $t {\tiny \begin{array}{l}  >\\ \sim \end{array}} 3$), 
the universal contribution to the $m_0^2$ term becomes small when 
$D_0 \approx 1/3$.

  \begin{figure}[t]
\begin{center}
\mbox{\psfig{figure=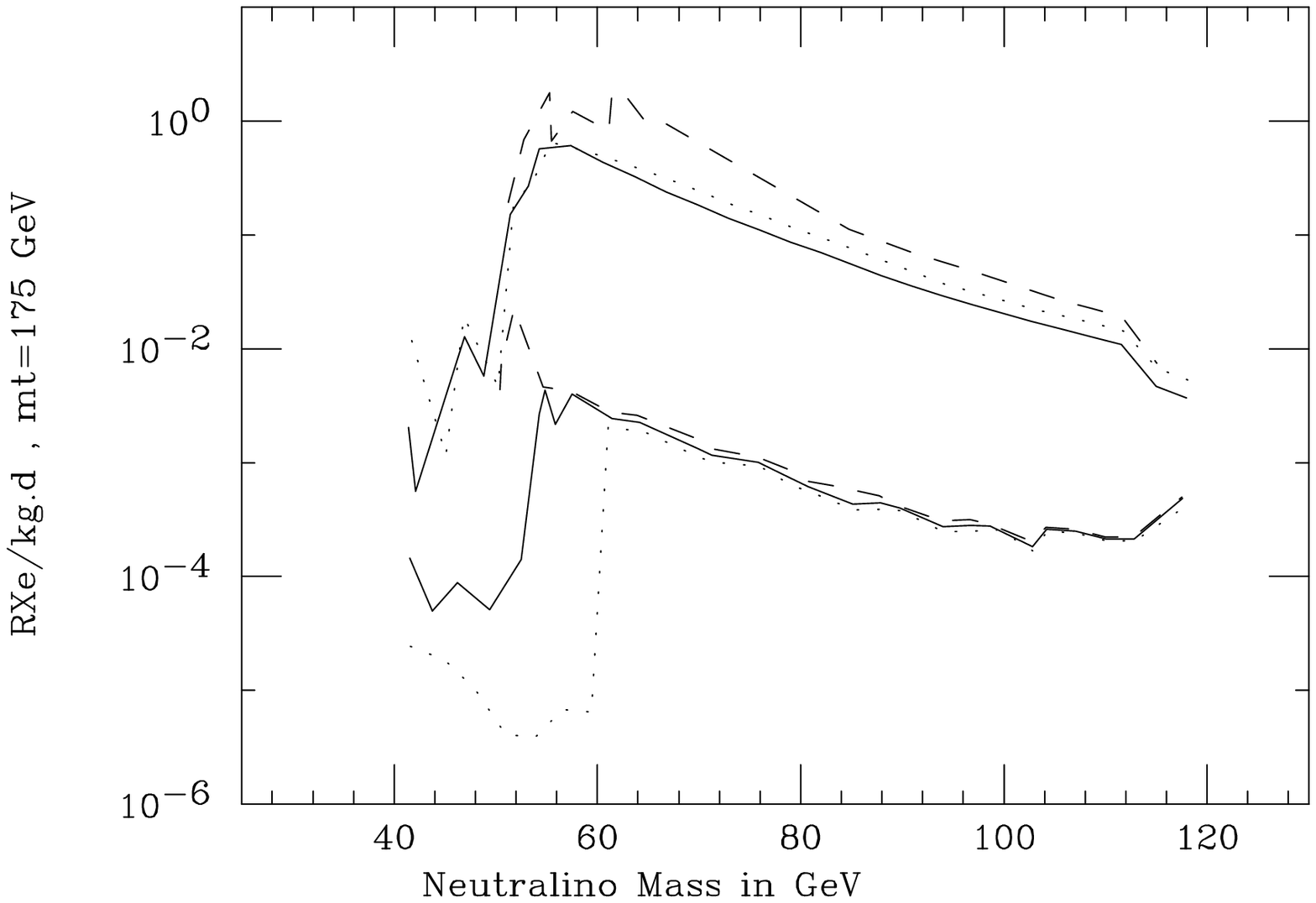,width=3in}}
\end{center}
 {\footnotesize 
Figure 1: 
 Maximum and minimum event rates for a Xe detector  as a function of 
 $m_{\chi_1^0}$ for $\mu>0$, $0.1\le \Omega_{\chi_1^0} h^2 \le 0.4$ with  
 $\delta_3=0=\delta_4$ and $\delta_2=0=\delta_1$ (solid), 
 $\delta_2=-1=-\delta_1$ (dotted),  $\delta_2=1=-\delta_1$ (dashed) \cite{13}.
  } \end{figure}

The above ideas allow one to understand qualitatively the detailed computer 
calculations of event rates \cite{13}. 
Fig.1 shows maximum and minimum event rates for the case 
$\delta_3=0=\delta_4$. 
The solid curves are for universal soft breaking.
From Eq.(17), the case $\delta_2=-1=-\delta_1$ (dotted curves) corresponds to 
increasing $\mu^2$, which reduces the event rates. 
This reduction can be as much as a factor of $O(10-100)$ for 
$m_{\chi_1^0} {\tiny \begin{array}{l}  <\\ \sim \end{array}} 60$ $GeV$ .
The   effect is largest for the minimum event rates, since these occur at
small $\tan \beta$ (which by Eq.(17) would magnify the effect). 
The reverse situation,  $\delta_2=1=-\delta_1$ (dashed curve) causes 
a decrease of 
$\mu^2$ and hence increase of event rates.
The effects of non-universal soft breaking masses becomes small for heavy neutralinos, 
$m_{\chi_1^0} {\tiny \begin{array}{l}  >\\ \sim \end{array}} 60$ $GeV$,
since then the $C_{\mu} m_{1/2}$ becomes larger masking the non-universal
effects.

 \begin{figure}
\begin{center}
\mbox{\psfig{figure=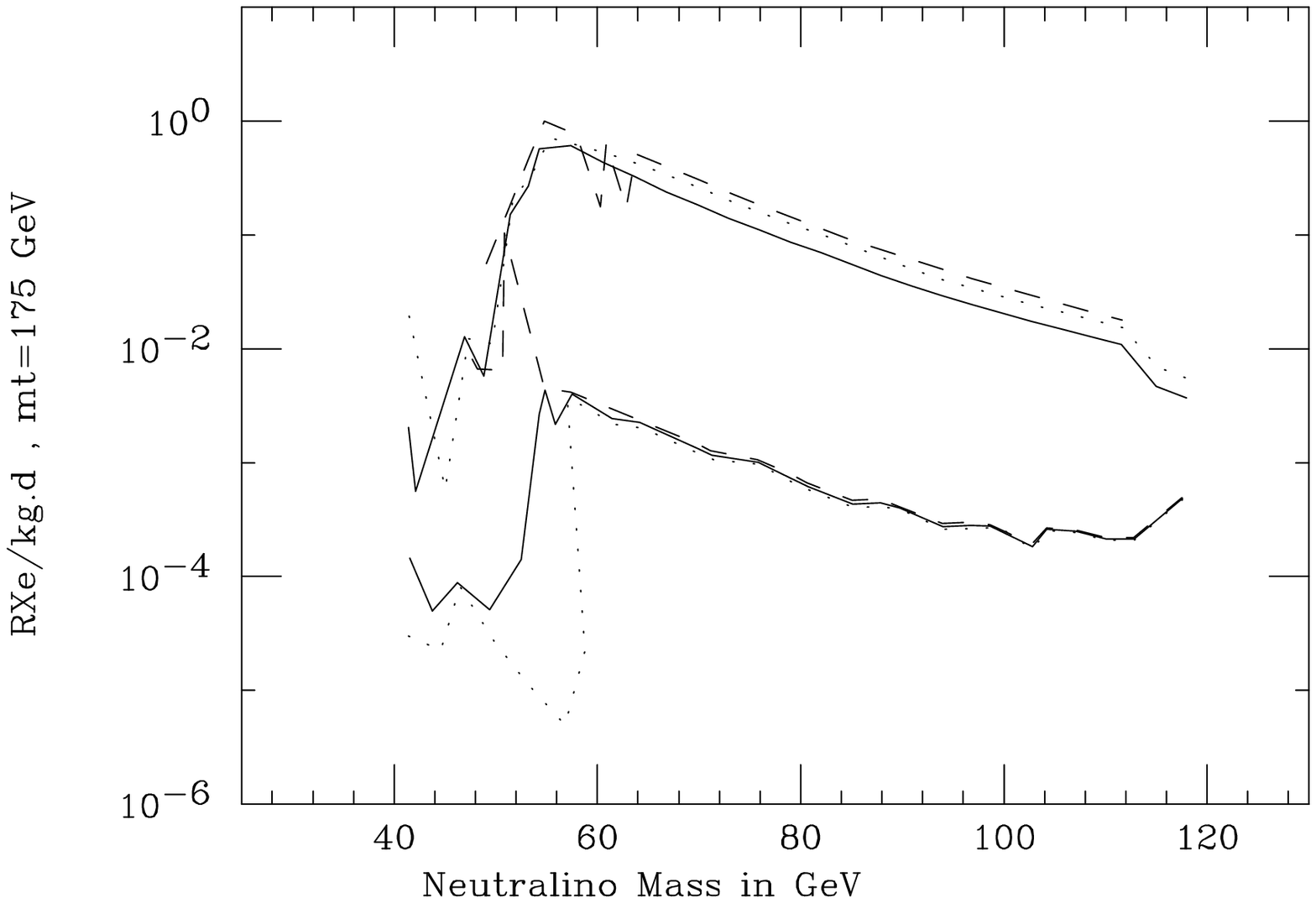,width=3in}}
\end{center}
{\footnotesize 
 Figure 2: Same as Fig.1 with $\delta_1=\delta_2=\delta_3=0$ and $\delta_4=0$
 (solid), $\delta_4=+1$ (dotted), $\delta=-1$ (dashed) \cite{13}. 
  } \end{figure}

Fig.2 shows a similar analysis with $\delta_1$, $\delta_2$ and $\delta_3$ 
set to zero. 
We see the curves with $\delta_4=1$ (dotted) resembles the $\delta_2=-1=-\delta_1$
curves of Fig.1 while the $\delta_4=-1$ (dashed) resembles the 
$\delta_2=1=-\delta_1$ curves of Fig.1. 
Fig.3 shows curves similar to Fig.2 with $\delta_3$ non-zero.
These results follow from Eq.(17) where the effects of $\delta_3$ and $\delta_4$
were seen to be opposite to those of $\delta_2$, e.g. a negative $\delta_4$
simulating a positive $\delta_2$ etc.

\begin{figure}
\begin{center}
\mbox{\psfig{figure=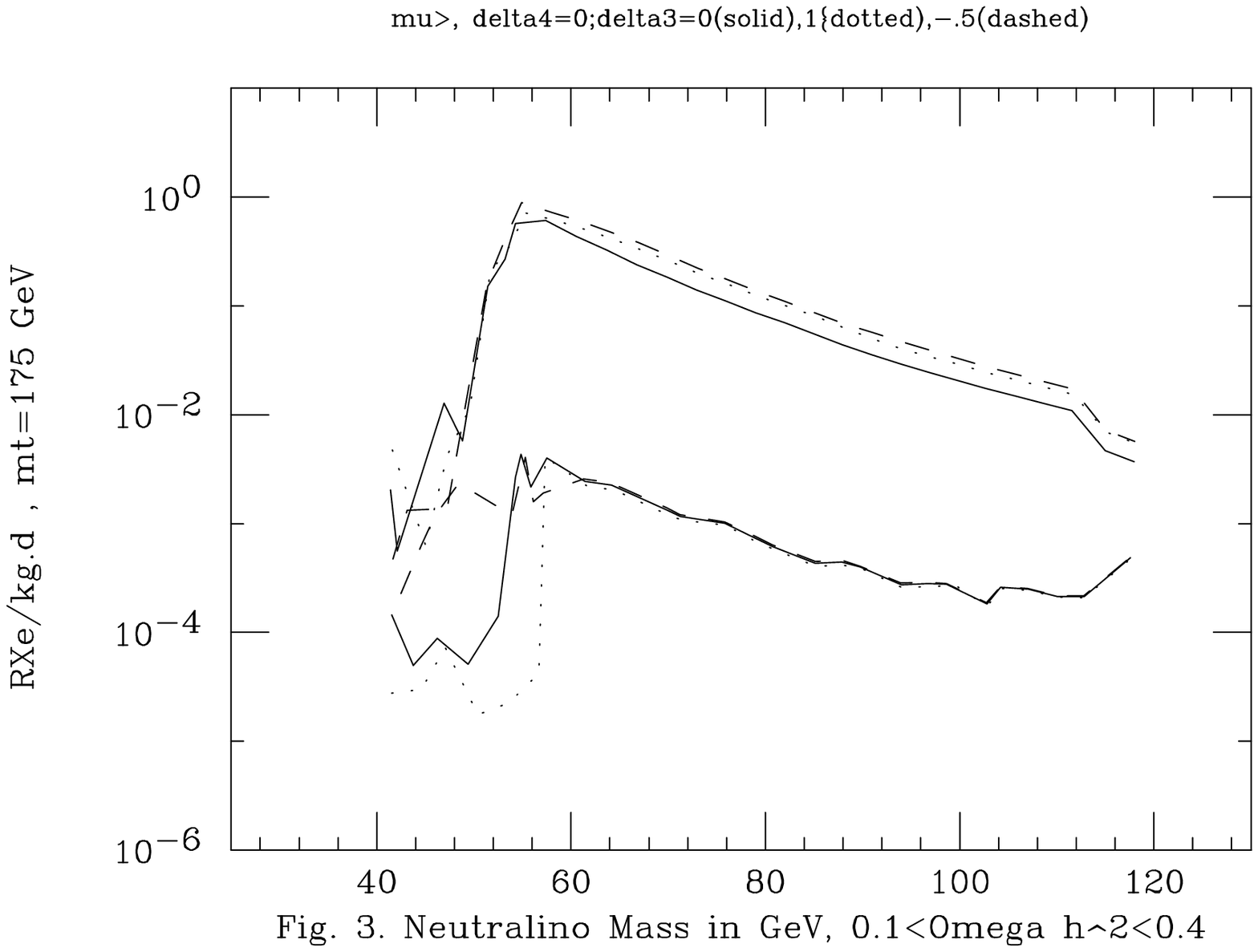,width=3in}}
\end{center}
{\footnotesize 
Figure 3: Same as Fig.1 with $\delta_1=\delta_2=\delta_4=0$ and 
$\delta_3=0$ (solid), $\delta_3= +1$ (dotted), 
$\delta_3=-1$ (dashed). \cite{13} 
} \end{figure}

\begin{figure}
\begin{center}
\mbox{\psfig{figure=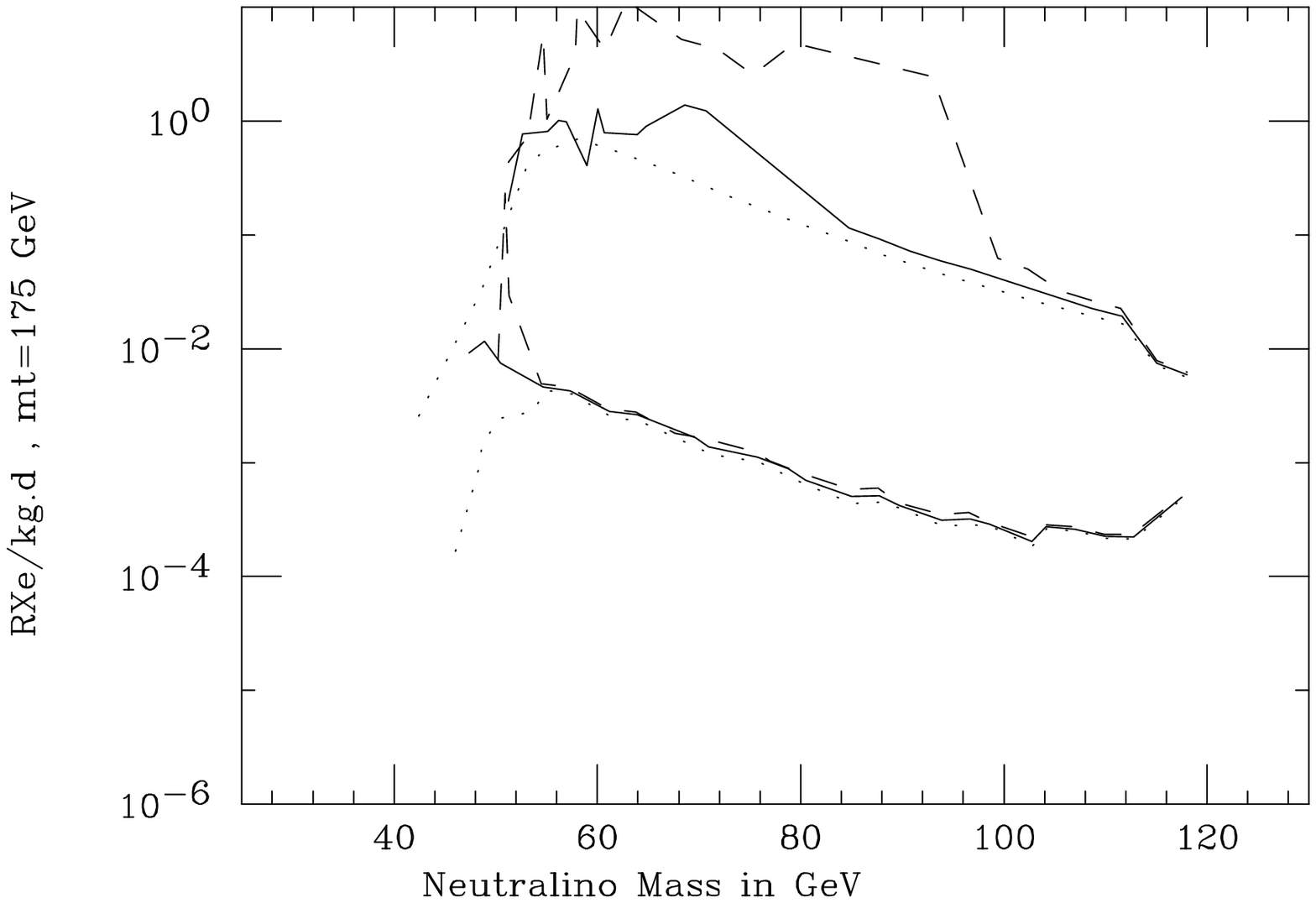,width=3in}}
\end{center}
 { \footnotesize 
 Figure 4:
 Same as Fig.1 with $\delta_3=-1=\delta_4$ and  $\delta_2=0=\delta_1$
(solid), $\delta_2=-1=-\delta_1$ (dotted), $\delta_2=1=-\delta_1$ 
(dashed ).\cite{13} 
  } \end{figure}

As seen in Eq.(16), $\delta_3=\delta_4$ for most GUT groups.
Fig.4 examines this case with the choice $\delta_3=\delta_4=-1$.
Here the squark non-universal masses add coherently in Eq.(17) with the Higgs 
non-universal masses for the case  $\delta_2=1=-\delta_1$ to reduce $\mu^2$.
The maximum event rate for this case (dashed curve) which occurs for large 
$\tan\beta$, is then greatly enhanced over a wide range of $m_{\chi_1^0}$ 
rising to $R\sim(1-10)$ $events/kg$ $d$. 
This value is the current sensitivity of the NaI detectors \cite{14}, and so this case 
is at the edge of experimental observation.
For the reverse situation, $\delta_2=-1=-\delta_1$ (dotted curve), the squark 
and Higgs non-universal effects mostly cancel in Eq.(17) and one indeed finds 
that these curves lie close to the universal one for most of the neutralino mass 
range, even though a large amount of non-universal soft breaking has occured.

The above discussion shows how squark and Higgs non-universal soft breaking 
masses can interact with each other, sometimes enhancing and sometimes canceling 
their effects depending on the sign of the non-universal terms.
 
\section{The $t-$Quark Mass, $b\rightarrow s+\gamma$ 
Decay And The Sign Of $\mu$}

While the value of $\mu^2$ is a crucial parameter in determining DM event rates, 
non-universal soft breaking masses also enter in other parameters which control 
the event rates in a more indirect way. 
Thus the stop mass matrix is given by  
\begin{equation}
  M_{\tilde{t}}^2 = \pmatrix{  m_{t_L}^2    &  -m_t(A_t+\mu\cot\beta)  \cr
                                      -m_t(A_t+\mu\cot\beta)      & m_{t_R}^2}.
\end{equation}
where 
\begin{equation}
m_{t_L}^2 = m_{Q_L}^2 + m_t^2 + 
({1\over2 }- {2\over 3}\sin^2\Theta_W) M_Z^2 \cos 2\beta
\end{equation}
\begin{equation}
m_{t_R}^2 = m_U^2 + m_t^2 + 
 {2\over 3}\sin^2\Theta_W M_Z^2 \cos 2\beta
\end{equation}
Here \cite{13}
\bea
m_{Q_L}^2 & = &  \biggl[ \biggl \{ {1+D_0\over 2} \biggr \}
       + \biggl \{ {5+D_0\over 6}\delta_3
      - {1-D_0 \over 6} (\delta_2+\delta_4)  \biggr \} \biggr ] m_0^2 
\nonumber \\
 &  &  - {1\over 6}(1-D_0){A_R^2\over D_0} + C_Qm_{1/2}^2
    - {1\over 66}(1-{\alpha_1(M_Z)\over \alpha_G}) S_0
 \eea
\bea
m_U^2 & = &  \biggl[\{ D_0\} 
        + \biggl \{ {2+D_0\over 3}\delta_4
      - {1-D_0 \over 3} (\delta_2+\delta_3)  \biggr \} \biggr ] m_0^2 
\nonumber \\
 &  &  - {1\over 3}(1-D_0){A_R^2\over D_0} + C_Um_{1/2}^2
    + {2\over 33}(1-{\alpha_1(M_Z)\over \alpha_G}) S_0
 \eea
and $C_Q$, $C_U$ are given in Ref.\cite{13}.
Note here $\delta_3$ and $\delta_4$ act oppositely to each other, though 
if the GUT condition $\delta_3=\delta_4$ is imposed, they still act oppositely 
to  $\delta_2$. 
 
Since the top quark is heavy, i.e. $D_0$ is small, the Landau pole term in 
Eq.(22) can drive $\tilde{t}_1$, the light stop, tachyonic if $A_R$ is large enough, 
eliminating such parameter points \cite{15}. 
Since  $A_R^2\cong (A_t-0.61m_{\tilde{g}})^2$, regions where $A_t$ is 
negative get eliminated, and one finds the allowed region of parameter 
space is restricted  by \cite{15,16}
\begin{equation}
A_t/m_0 {\tiny \begin{array}{l}  >\\ \sim \end{array}} -0.5
\end{equation}

The $b\rightarrow s+\gamma$ decay also strongly restricts the SUSY parameter 
\break space \cite{17,18}. 
The measured branching ratio for this process by CLEO is \cite{19}  
$B(B\rightarrow X_s\gamma) = (2.32\pm 0.67)\times 10^{-4}$ .
The SM prediction, with NLO effects included is 
$B(B\rightarrow X_s\gamma) = (3.48\pm 0.31)\times 10^{-4}$ where 
the error in the theoretical calculations is now dominated by uncertainties 
in the input parameters, e.g. $\alpha_3(M_Z)$, $m_t$, $m_c/m_b$, etc.\cite{20} 
While these two results are statistically consistent, it is clear that there is 
a certain amount of tension between the data and the SM.
The SUSY corrections to the SM reduce the SM branching ratio for $A_t/\mu>0$ 
and increase it or $A_t/\mu<0$. 
Since by Eq.(23) $A_t$ is mostly positive, and the experimental branching ratio 
lies below the SM value, the current CLEO data already eliminates most 
of the $\mu<0$ part of the parameter space at the 95\% C.L. level.
The part of the $\mu<0$ parameter space that survives is for small $\tan\beta$, 
and so one finds in most of the remaining parameter space, the event rates for 
$\mu<0$  are a factor of 100 (or more) smaller than those of $\mu>0$ and this 
result is true with or without non-universal soft breaking masses \cite{13}. 

A second feature involving the $b\rightarrow s+\gamma$ decay is the existance 
of a correlation between  the DM event rate $R$ and the $b\rightarrow s+\gamma$ 
branching ratio $B$: large (small) event rates tend to correlate with small (large) 
branching ratio.
This shows up most strongly if one considers the maximim $(R_{max})$ or 
minimum $(R_{min})$ event rate at fixed values of the neutralino mass. 
One finds (for universal soft breaking) that the part of the parameter space 
with $B\le 2\times 10^{-4}$ corresponds to 
$R_{max} {\tiny \begin{array}{l}  >\\ \sim \end{array}} 0.1 \ events/kg$ $d$, and 
$B\ge 3\times 10^{-4}$ corresponds to 
$R_{max} {\tiny \begin{array}{l}  <\\ \sim \end{array}} 0.01\ events/kg\ d$. 
Similarly for $R_{min}\le 0.003\ events/kg\ d$, one finds $B>3\times 10^{-4}$. 
The correlation is less strong for values of $R$ in between $R_{max}$ and 
$R_{min}$,  but still significant \cite{21}.
  
It is clear from the above that as the $b\rightarrow s+\gamma$ data becomes 
more precise, it will have a strong impact on DM predictions, and on restricting 
the supersymmetry parameter space in general.

\section{Determination Of Cosmological Parameters}

One of the major sources of uncertainty in dark matter event rate predictions 
is the lack of accurate knowledge of the basic cosmological parameters.
Future satellite experiments \cite{22}, MAP (scheduled for the year 2000) and 
PLANCK (planned for 2005) will be able to measure the angular power spectrum 
 very accurately. 
Thus one defines the correlation function 
\begin{equation}
c(\theta) = \langle {\Delta T(\hat{q}_1)\over T_0} 
{\Delta T(\hat{q}_2)\over T_0}\rangle 
\end{equation}
where $T_0=2.728\pm0.002^{\circ}K$ is the cosmic microwave background 
(CMB) temperature, $\Delta T(\hat{q}_i)$, $i=1,2$ are the deviations from $T_0$ 
in directions $\hat{q}_i$ and $\cos\theta= \hat{q}_1 \cdot \hat{q}_2$.
Expanding in Legendre polynomials
\begin{equation}
c(\theta)= \sum {2l+1\over 4\pi} c_lP_l(\cos\theta)
\end{equation}  
these experiments will be able to determine the $c_l$ out to $l\approx 1000$. 
Since the $c_l$ are sensitive to the different cosmological parameters, it will 
be possible to determine the Hubble constant, the amount of dark matter, the 
cosmological constant etc. to an accuracy of a few percent \cite{23,24}. 
We consider here what such determinations might mean for DM predictions 
for two cosmological models.

\subsection{$\Lambda$CDM Model}

Current astronomical measurements suggest that while there is a large amount 
of CDM,  it's mean density is considerably less than the critical density $\rho_c$ 
and that models with a cosmological constant $\Lambda$ and 
$\Omega_{CDM}\approx 0.4$ are good, if not better fits to the current 
astronomical data \cite{25}.  
We consider here the  $\Lambda$CDM model with cold and baryonic DM such 
that $\Omega_{total}=1$ where 
$\Omega_{total}=\Omega_{CDM}+\Omega_B+\Omega_\Lambda$. 
We assume that the central values of the cosmological parameters measured 
by PLANCK and MAP are 
\begin{equation}
\Omega_{CDM}=0.4, \quad \Omega_B=0.05, 
\quad \Omega_\Lambda=0.55, \quad h=0.62
\end{equation}
(consistent with current determinations of these quantities). 
The accuracy with which each of these quantities can be measured by the PLANCK 
satellites has been estimated in Ref.\cite{23} and one finds from this that \cite{21} 
\begin{equation}
\Omega_{CDM}h^2= 0.154\pm0.017
\end{equation} 
This is to be compared with the very broad window of Eq.(2) for the current estimates 
of $\Omega_{CDM}h^2$.

Assuming as before that the CDM is the relic neutralinos, this sharpening of 
the allowed region of $\Omega_{\chi_1^0}h^2$ significantly effects the DM event 
rate predictions. 
Fig.5 shows the maximum and minimum event rates as a function of 
$m_{\tilde{g}}$ for universal soft breaking ($\delta_i=0$). 
(Recall $m_{\chi_1^0}$ scales with $m_{\tilde{g}}$ by Eq.(10).) 
Comparing with e.g. the solid curve of Fig.1, we see that the effect of 
Eq.(27)

\begin{figure}
\begin{center}
\mbox{\psfig{figure=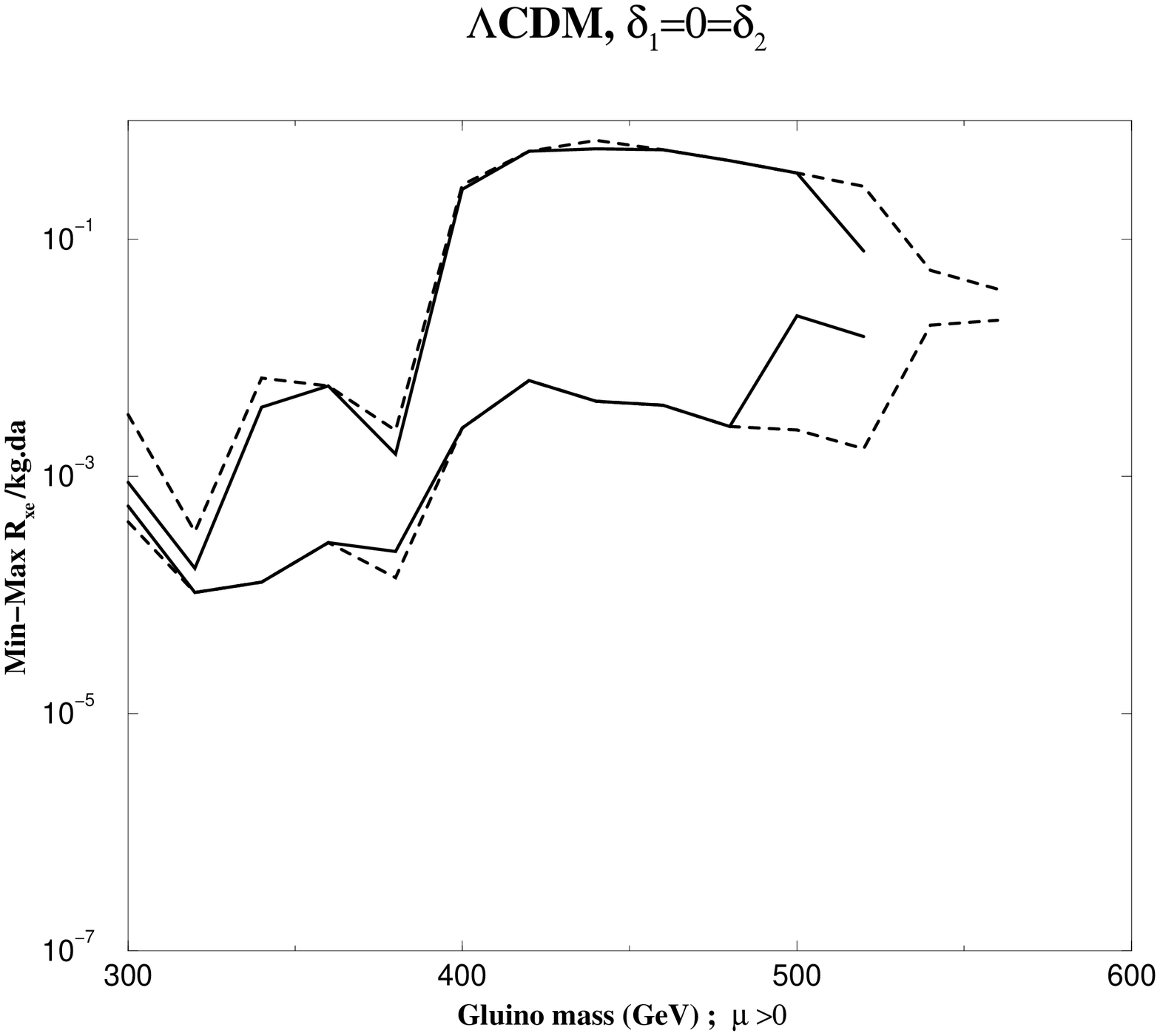,angle=270,width=3in}}
\end{center}
 {\footnotesize 
Figure 5: Maximum and minimum event rates for a Xe detector  for $\mu>0$ 
as a function of $m_{\tilde{g}}$ for universal soft breaking. 
The solid (dashed) curves are the 1std (2std) range of Eq.(27) \cite{21}. 
  } \end{figure}

\begin{figure}
\begin{center}
\mbox{\psfig{figure=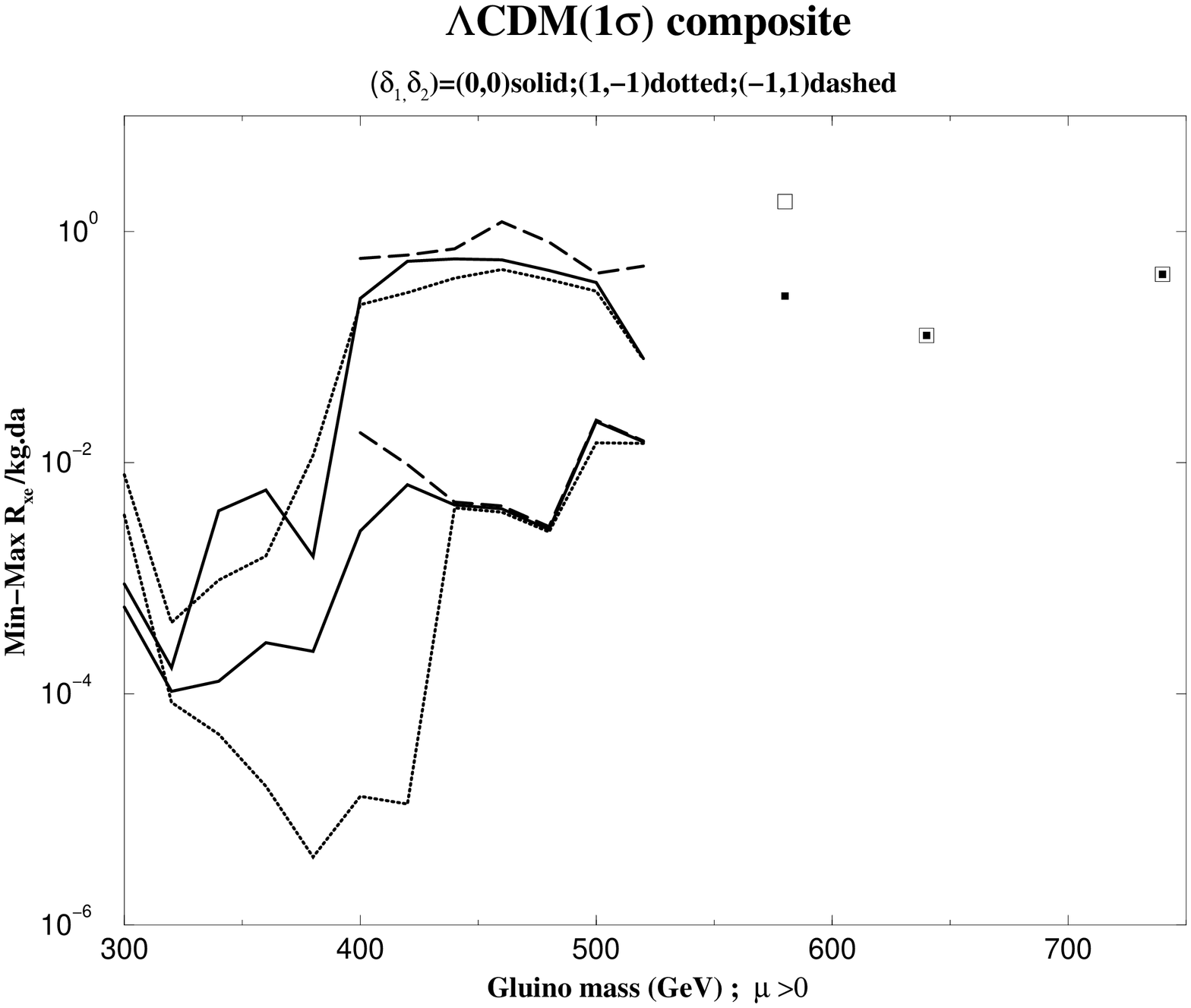,width=3in,angle=270}}
\end{center}
 {\footnotesize 
Figure 6:  Same as Fig.5 for 1 std range of Eq.(27) with  $\delta_2=0=\delta_1$ 
(solid), $\delta_2=-1=-\delta_1$ (dotted), $\delta_2=1=-\delta_1$ (dashed) 
and $\delta_3=0=\delta_4$. 
 The isolated points are for $\delta_2=1=-\delta_1$ ($R_{max}$ are empty 
 squares, $R_{min}$ are solid squares).\cite{21}
  }
\end{figure}

\noindent
is to increase $R_{min}$ significantly and put an upper bound on 
$m_{\tilde{g}}$ (and hence on   $m_{\chi_1^0}$ ). 
This latter effect arises from the fact that the  $\chi_1^0$ annihilation cross 
section in the early universe is a decreasing function of $m_{\chi_1^0}$, and 
hence an upper bound on the relic density produces an upper bound on 
$m_{\chi_1^0}$. 
Fig.6 shows the effects of non-universal soft breaking on the event rates. 
Note that for $\delta_2=1=-\delta_1$ there is also a minimum value of 
$m_{\tilde{g}}$ (and hence a minimum $m_{\chi_1^0}$) since for this case 
a lower value of $m_{\chi_1^0}$ would violate the lower bound on 
$\Omega_{\chi_1^0}h^2$ of Eq.(27).
(The additional isolated points for $\delta_2=1=-\delta_1$ arise from the fact 
that for this case $\mu^2$ is driven small (as can be seen from Eq.(17)) and so 
the scaling relations Eq.(10) no longer hold allowing $m_{\chi_1^0}$ to be 
sufficiently small so that the upper bound of Eq.(27) can be satisfied even though 
$m_{\tilde{g}}$ is large.) 
We also note that for $\delta_2=1=-\delta_1$, almost the entire parameter space 
has now been eliminated for $\mu<0$. 

The above discussion shows that the upper bound on $\Omega_{\chi_1^0}h^2$  of 
Eq.(27) for the $\Lambda$CDM model produces upper bounds on $m_{\tilde{g}}$ 
and by scaling, on the other gauginos. 
Thus (aside from the discrete points in Fig.6) the 1std (2std) gaugino upper bounds 
are $m_{\tilde{g}}\le 520(560)$ $GeV$, $m_{\chi_1^0}\le 70(77)$ $GeV$, 
$m_{\chi_1^\pm}  {\tiny \begin{array}{l}  <\\ \sim \end{array}} 150$ $GeV$,
and also the light Higgs obeys 
$m_h {\tiny \begin{array}{l}  <\\ \sim \end{array}} 120$ $GeV$. 
It is interesting to compare these bounds with what might be the expected 
reach of an upgraded Tevatron with $25fb^{-1}$ of data. 
Thus it is estimated that the gluino could be observed with a mass reach of 
$ m_{\tilde{g}} {\tiny \begin{array}{l}  <\\ \sim \end{array}} 450$ $GeV$ 
for most of the parameter space, the chargino with 
 $ m_{\chi_1^\pm}  {\tiny \begin{array}{l}  <\\ \sim \end{array}} 150$ $GeV$ 
for about  $2/3$ of the parameter space and a Higgs for   
$m_h {\tiny \begin{array}{l}  <\\ \sim \end{array}} 120$ $GeV$ \cite{26}.
Hence the example of the $\Lambda$CDM model considered here would 
suggest that an upgraded Tevatron could examine a significant part of the 
SUSY parameter space allowed by the PLANCK satellite measurements 
of the cosmological parameters. 

As discussed in Sec.6, there is a correlation between large (small) DM 
event rates $R$ and small (large) $b\rightarrow s+\gamma$ branching ratios $B$. 
This is exhibited in detail for the $\Lambda$CDM model in Fig.7 which shows a 
general scatter plot over the full parameter space of the model of Eq.(27). 
Recall that the curent CLEO data is \cite{19} $B=(2.32\pm 0.67)\times 10^{-4}$ 
while the SM prediction is  \cite{20} $B=(3.48\pm 0.31)\times 10^{-4}$.
One sees from Fig.7 that if $B>3\times 10^{-4}$ (consistent with the SM 
predictions) then $R$ is small i.e. $R<0.1\ events/kg\ d$ while if $B<3\times 10^{-4}$ 
(suggested by the current CLEO data) then $R>0.05\ events/kg\ d$ 
for almost 
all parameter points. 
Thus one finds an interesting correlation between accelerator physics and 
cosmology.

\begin{figure}
\begin{center}
\mbox{\psfig{figure=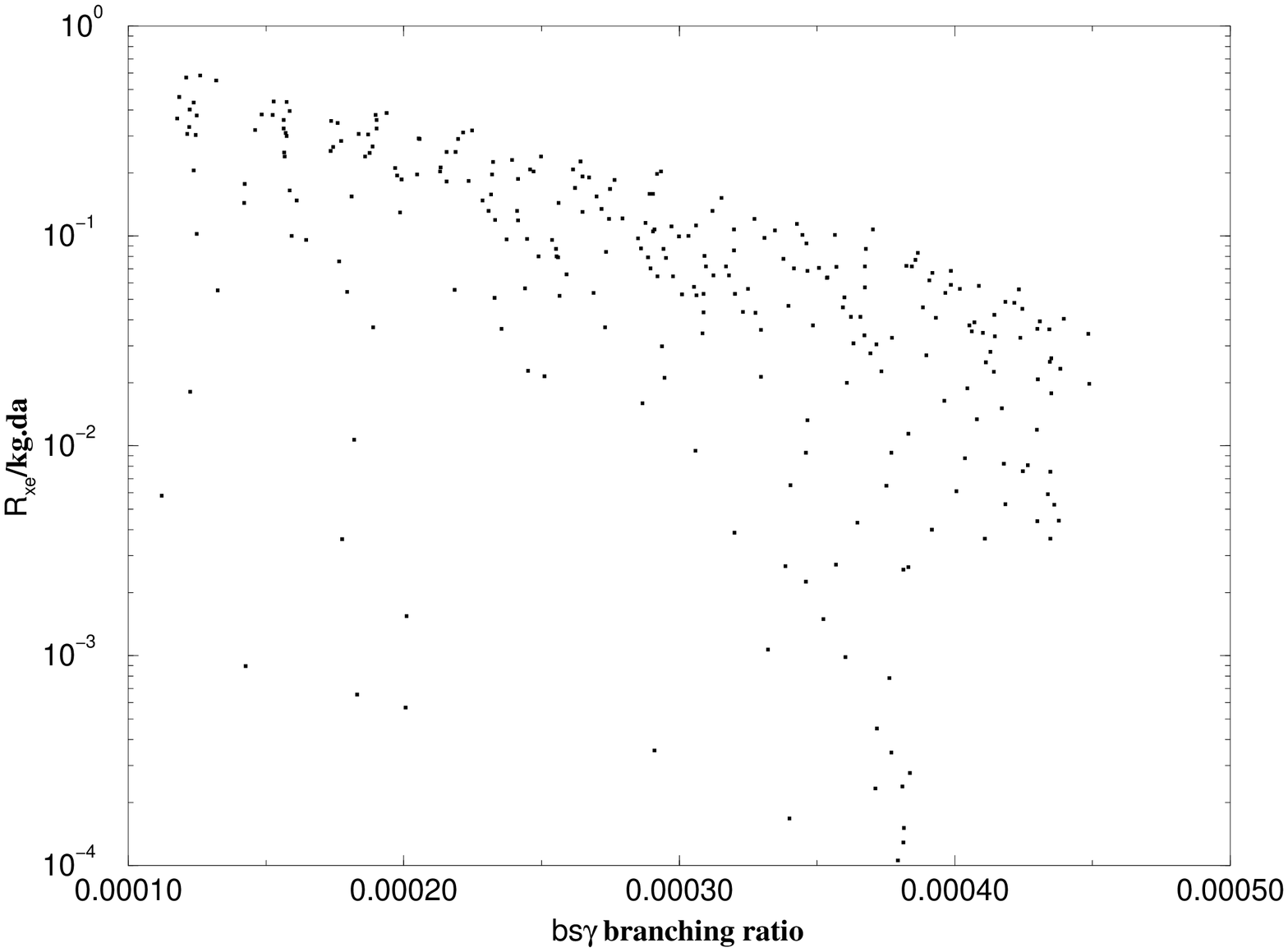,width=3in}}
\end{center}
 {\footnotesize 
Figure 7: Scatter plot over allowed SUSY parameter space for R vs. B 
($b\rightarrow s+\gamma$) for $\Lambda$CDM model for $\mu>0$ with 
universal soft breaking masses and 1std bounds of Eq.(27). \cite{21} 
  } \end{figure}

\subsection{$\nu$CDM model}

If neutrinos do have masses in the $eV$ range, they would contribute significantly 
to the hot dark matter of the universe. 
As a second example of what the new satellite experiments might see, we consider 
the neutrino cold dark matter ($\nu$CDM) model and assume that the MAP and 
PLANCK  satellites determine the following central values for the cosmological 
parameters:
\begin{equation}
\Omega_\nu=0.2; \quad \Omega_{CDM}=0.75; 
\quad \Omega_B=0.05; \quad h=0.62
\end{equation}
Using the estimated accuracies that these quantities can be measured by the 
PLANCK satellite \cite{23,24} one finds \cite{21}
\begin{equation}
\Omega_{CDM}h^2 = 0.288 \pm 0.013 
\end{equation}
As in the $\Lambda$CDM model, one finds a narrowing in the difference between 
the maximum and minimum event rates for 
$m_{\chi_1^0} {\tiny \begin{array}{l}  <\\ \sim \end{array}} 60$ $GeV$. 
There is also an upper bound on the gaugino masses, which now is higher 
than in the $\Lambda$CDM model since $\Omega_{\chi_1^0}h^2$ is larger. 
One finds for the 1std (2std) bounds of Eq.(29) that \cite{21} 
$m_{\tilde{g}} {\tiny \begin{array}{l}  <\\ \sim \end{array}}  700(720)$ $GeV$, 
$m_{\chi_1^0} {\tiny \begin{array}{l}  <\\ \sim \end{array}}  95(100)$ $GeV$ and 
$m_{\chi_1^\pm} {\tiny \begin{array}{l}  <\\ \sim \end{array}} 200$ $GeV$.
Thus for this model, the LHC would be crucial in order to explore the full 
SUSY parameter space. 
The 1std window of Eq.(29) also shows an additional feature of gaps in the 
allowed range of $m_{\tilde{g}}$ (and $m_{\chi_1^0}$). 
While these gaps fill in at the 2std level, they illustrate the impact that 
astronomical  measurements may have on accelerator physics searches.  

\section{Conclusions}

We have considered here the effects of non-universal SUSY soft breaking 
on predictions of dark matter detection rates for supergravity models with 
$R$ parity where supersymmetry is broken at a scale 
${\tiny \begin{array}{l}  >\\ \sim \end{array}} M_G$, 
the breaking being communicated by gravity to the physical sector. 
The lightest neutralino becomes the CDM candidate for almost the entire 
parameter space.
For models which have universal soft breaking in the first two generations 
(to suppress FCNC processes), one generally needs nine additional parameters 
to describe the non-universal effects of the Higgs boson and third generation 
squark and slepton masses (which reduce to five parameters if one imposes the symmetry 
constraints, Eq.(16), of most GUT groups). 
For $\tan\beta {\tiny \begin{array}{l}  <\\ \sim \end{array}} 20$ only four of these, 
two Higgs masses and two squark masses, enter significantly. 
While this is still a major enlargement of the minimal SUSY parameter space, 
the extensive knowledge already obtained from studying SUGRA dark matter 
predictions for universal soft breaking have allowed one to determine in large 
measure the effects that such non-universal soft breaking produce. 
Thus the squark and Higgs non-universal effects can act to 
enhance or cancel each other (depending on their relative signs) and can increase 
or decrease event rates by a factor of $\approx 10-100$ in the domain 
$m_{\chi_1^0} {\tiny \begin{array}{l}  <\\ \sim \end{array}} 60$ $GeV$ 
($m_{\tilde{g}} {\tiny \begin{array}{l}  <\\ \sim \end{array}} 400$ $GeV$ ). 
However, the non-universal effects are generally small at higher masses \cite{13}. 

SUGRA models can account for both the existance of CDM as well as accelerator 
SUSY phenomena, and interaction between the two sets of phenomena has begun to 
occur. 
Thus the fact that the top quark is heavy, and that the measured 
$b\rightarrow s+\gamma$ branching ratio lies about 1.6 std below the SM prediction 
eliminates most of the $\mu<0$ and $A_t<0$ parts of the parameter space at the 
95\% C.L. $^{15-18}$ 
This then leads to DM event rate predictions for the small remaining part of 
the parameter space with $\mu<0$ to be $\approx 100$ times 
smaller than for $\mu>0$ \cite{13}. 
Further one finds a correlation between large (small) event rates and small (large) 
predicted   $b\rightarrow s+\gamma$ branching ratio for SUGRA models \cite{21}. 
One expects that further $b\rightarrow s+\gamma$ data will play an important role in SUGRA 
dark matter event rate predictions. 

Future satellite experiments by PLANCK and MAP as well as ballon and ground based 
experiments will be able to determine the basic cosmological parameters with great 
accuracy (at the 1-10\% level) and hence greatly reduce the uncertainties in DM 
predictions. 
To illustrate what might be expected from the PLANCK and MAP experiments, 
examples of the $\Lambda$CDM model and $\nu$CDM model were considered. 
The narrowing of the allowed window for $\Omega_{\chi_1^0}h^2$ generally 
reduces the spread between the maximum and minimum event rates for  
$m_{\chi_1^0} {\tiny \begin{array}{l}  <\\ \sim \end{array}} 60$ $GeV$, 
and limits the maximum  (and for some signs of non-universal soft breaking 
the minimum) values of gaugino masses \cite{21}. 
Thus astronomical measurements can have a significant impact on SUSY 
accelerator searches. 

\section*{Acknowledgement}

This work was supported in part by a National Science Foundation Grant 
PHY-9722090.


\section*{References}
 
\begin{thebibliography}{99}
 
 \bibitem{jed} For a review see G. Jungman, M. Kamionkowski and K. Greist, 
{\em Supersymetric Dark Matter}, 
{\em Phys. Rep.} {\bf 267}, 195 (1995). 
 
 \bibitem{dva} A.H. Chamseddine, R. Arnowitt and P. Nath, 
{\em Phys. Rev. Lett.} {\bf 49}, 970 (1982). 
For reviews see P. Nath,  R. Arnowitt and A.H. Chamseddine,
{\em Applied N=1 Supergravity} 
(World Scientific, Singapore, 1984);
H.P. Nilles, {\em Phys. Rep.} {\bf 110}, 1 (1984);  
R. Arnowitt and P. Nath, Proc. of VII J.A. Swieca Summer School ed. 
E. Eboli (World Scientific, Singapore, 1994).
 
 \bibitem{tri} E.W. Kolb and M.S. Turner, 
{\em The Early Universe} (Addison-Wesley, Redwood City, 1989).

 \bibitem{cet} L. Hall, J. Lykken and S. Weinberg, 
{\em Phys. Rev.} D {\bf 27}, 2359 (1983);
P. Nath,  R. Arnowitt and A.H. Chamseddine,
{\em Nucl. Phys.} B {\bf 227}, 121 (1983).
 
 \bibitem{pet}K. Inoue et al. {\em Prog. Theor. Phys.} {\bf 68}, 927 (1982);
L. Iba\~nez and G.G. Ross, 
{\em Phys. Lett.} B {\bf 110}, 227 (1982);
L. Alvarez-Gaum\'e, J. Polchinski and M.B. Wise, 
{\em Nucl. Phys.} B {\bf 221}, 495 (1983);
J. Ellis, J. Hagelin, D.V. Nanopoulos and K. Tamvakis, 
{\em Phys. Lett.} B {\bf 125}, 2275 (1983); 
L.E. Iba\~nez and C. Lopez, 
{\em Nucl. Phys.} B {\bf 233}, 545 (1984);
L.E. Iba\~nez, C. Lopez and C. Mu\~nos, 
{\em Nucl. Phys.} B {\bf 256}, 218 (1985).
 
 \bibitem{ses} J. Ellis and F. Zwirner, 
{\em Nucl. Phys.} B {\bf 338}, 317 (1990);
R. Arnowitt and P. Nath, 
{\em Phys. Rev.} D {\bf 46}, 3981 (1992).

 \bibitem{sed} R. Arnowitt and P. Nath, 
{\em Phys. Rev. Lett.} {\bf 69}, 725 (1992). 
 P. Nath and R. Arnowitt, 
{\em Phys. Lett.} B {\bf 289}, 368 (1992). 
 
 \bibitem{osa} For recent papers see 
R. Arnowitt and P. Nath, {\em Mod. Phys. Lett.} A {\bf 10}, 1275 (1995);
P. Nath and R. Arnowitt, {\em Phys. Rev. Lett.} {\bf 74}, 4592 (1995);
E. Diehl, G. Kane C. Kolda and J. Wells,  {\em Phys. Rev.} D {\bf 52}, 4223 (1995); 
 R. Arnowitt and P. Nath, {\em Phys. Rev.} D {\bf 54}, 2374 (1996); 
L. Bergstrom and P. Gondolo, {\em Astropart. Phys.} {\bf 5}, 263 (1996);
J.D. Vergados, {\em J. Phys.} G {\bf 22}, 253 (1996);
V.A. Bednyakov, S.G. Kovalenko, H.V. Klapdor-Kleingrothaus and Y. Ramachers, 
{\em Z. Phys.} A {\bf 357}, 339 (1997);
J. Edsj\"o and P. Gondolo, hep-ph/9704361;
V. Barger and C. Kao, hep-ph/9704403;
H. Baer and M. Brhlik, hep-ph/9706509. 
Analyses with non-universal Higgs soft breaking masses are given in 
D. Matalliotakis and H.P. Nilles, {\em Nucl. Phys.} B {\bf 435}, 115 (1995); 
M. Olechowski and S. Pokorski, {\em Phys. Lett.} B {\bf 344}, 201 (1995); 
V. Berezinsky, A. Bottino, J. Ellis, N. Forrengo, G. Mignola and S. Scopel, 
{\em Astropart. Phys.} {\bf 5}, 1 (1996).

 \bibitem{dev} S.K. Soni and H.A. Weldon, 
{\em Phys. Lett.} B {\bf 126}, 215 (1983);  
V.S. Kaplunovsky and J. Louis, 
{\em Phys. Lett.} B {\bf 306}, 268 (1993).
 
 \bibitem{des} N. Polonski and A. Pomerol, 
{\em Phys. Rev.} D {\bf 51}, 6532 (1995);
R. Barbieri, L. Hall and A. Strumia, 
{\em Nucl. Phys.} B {\bf 445}, 219 (1995).
 
 \bibitem{11} M. Drees, {\em Phys. Lett.} B {\bf 181}, 279 (1986); 
 P. Nath and R. Arnowitt, 
{\em Phys. Rev.} D {\bf 39}, 2006 (1989);
J.S. Hagelin and S. Kelley, 
{\em Nucl. Phys.} B {\bf 342}, 95 (1990);  
Y. Kawamura, H. Murayama and M. Yamaguchi, 
{\em Phys. Lett.} B {\bf 324}, 52 (1994).
 
\bibitem{12} S. Dimopoulos and H. Georgi, 
{\em Nucl. Phys.} B {\bf 206}, 387 (1981). 

\bibitem{13} P. Nath and R. Arnowitt, hep-ph/9701301.

\bibitem{14} A. Bottino, F. Donato, G. Mignola and S. Scopel,  P. Belli and 
A. Incicchitti,  {\em Phys. Lett.} B {\bf 402}, 113 (1997).

\bibitem{15} P. Nath, J. Wu and R. Arnowitt,
{\em Phys. Rev.} D {\bf 52}, 4169 (1995).

\bibitem{16} J. Wu, R. Arnowitt and P. Nath, 
{\em Phys. Rev.} D {\bf 51}, 1371 (1995).

\bibitem{17} P. Nath and R. Arnowitt,
{\em Phys. Lett.} B {\bf 336}, 395 (1994);
F. Borzumati, M. Drees and M. Nojiri, 
{\em Phys. Rev.} D {\bf 51}, 341 (1995).

\bibitem{18} P. Nath and R. Arnowitt,
{\em Phys. Rev. Lett.} {\bf 74}, 4592 (1995);
R. Arnowitt and P. Nath, 
{\em Phys. Rev.} D {\bf 54}, 2374 (1996);
H. Baer and M. Brhlik, 
{\em Phys. Rev.} D {\bf 55}, 3201 (1997).

\bibitem{19} M.S. Alam et al. (CLEO Collaboration), 
{\em Phys. Rev. Lett.} {\bf 74}, 2885 (1995). 

\bibitem{20} A.J. Buras, A. Kwiatkowski and N. Pott, 
hep-ph/9707482.  

\bibitem{21} R. Arnowitt and P. Nath, unpublished.

\bibitem{22} http://map.gsfc.nasa.gov/; 
http://astro.estec.esa.nl:80/SA-general/Projects/Cobras/cobras.html.

\bibitem{23} A. Kosowsky, M. Kamionkowski, G. Jungman and D. Spergel, 
{\em Nucl. Phys. Proc. Suppl.} {\bf 51B}, 49 (1996).

\bibitem{24} S. Dodelson, E. Gates and A. Stebbins, 
{\em Astroph. J.} {\bf 467}, 10 (1996). 

\bibitem{25} M.S. Turner, astro-ph/9703161; L. Krauss, astro-ph/9706227. 

\bibitem{26} T. Kamon, J.L. Lopez, P. McIntyre and J.T. White, 
{\em Phys. Rev.} D {\bf 50}, 5676 (1994); 
Report of the tev-2000 Study Group, eds. D. Amidei and R. Brock, 
FERMILAB-Pub-96/082. 

 \end{thebibliography}
 \end{document}